\documentclass[aps,prd,showpacs]{revtex4}

\usepackage{graphicx}
\usepackage{dcolumn}
\usepackage{bm}
\usepackage[dvips]{color}
\usepackage[normalem]{ulem}  
\usepackage{url}

\newcommand{\be}{\begin{eqnarray}} 
\newcommand{\ee}{\end{eqnarray}}

\begin{document}

\title{Tidal Love Numbers of Neutron and Self-Bound Quark Stars}
\author{Sergey Postnikov}
\affiliation{Department of Physics and Astrononmy, Ohio University, Athens,
  OH 45701-2979, USA}
\email{sp315503@ohio.edu, prakash@harsha.phy.ohiou.edu}
\author{Madappa Prakash}
\affiliation{Department of Physics and Astrononmy, Ohio University, Athens,
  OH 45701-2979, USA} \email{prakash@harsha.phy.ohiou.edu}
  \author{James M. Lattimer} \affiliation{Department of Physics and
  Astronomy, State University of New York at Stony Brook, Stony Brook,
  NY-11794-3800, USA} \email{lattimer@astro.sunysb.edu}
  
\date{\today}
  \begin{abstract}

  Gravitational waves from the final stages of inspiralling binary
  neutron stars are expected to be one of the most important sources
  for ground-based gravitational wave detectors.  The masses of the
  components are determinable from the orbital and chirp frequencies
  during the early part of the evolution, and large finite-size
  (tidal) effects are measurable toward the end of inspiral, but the
  gravitational wave signal is expected to be very complex at this
  time.  Tidal effects during the early part of the evolution will
  form a very small correction, but during this phase the signal is
  relatively clean.  The accumulated phase shift due to tidal
  corrections is characterized by a single quantity related to a
  star's tidal Love number.  The Love number is sensitive, in
  particular, to the compactness parameter $M/R$ and the star's
  internal structure, and its determination could provide an important
  constraint to the neutron star radius.  We show that the Love number
  of normal neutron stars are much different from those of self-bound
  strange quark matter stars.  Observations of the tidal signature
  from coalescing compact binaries could therefore provide an
  important, and possibly unique, way to distinguish self-bound
  strange quark stars from normal neutron stars.
  \end{abstract}

  \pacs{ 04.40.Dg, 26.60.Kp, 97.60.Jd, 95.85.Sz }
  
\maketitle

\section{INTRODUCTION}
Gravitational waves from the final stages of inspiralling binary
neutron stars are expected to be one of the most important sources for
ground-based gravitational wave detectors \citep{C1993}.  To date,
LIGO observations have only been able to set an upper limit to the
neutron star-neutron star coalescence rate of 0.039 yr$^{-1}
L_{10}^{-1}$ \citep{A2009}, where $L_{10}$ is the blue luminosity in
units of $10^{10} {\rm~L}_\odot$, which translates to about 0.075
events per year in the Milky Way.  This is a thousand times larger
than the predicted rates \citep{K2004}.  Nevertheless, the observed
neutron star-neutron star inspiral rate from the universe is expected
to be about 2 per day in LIGO II \citep{K2004}.  The masses of the
components will be determined to moderate accuracy, especially if the
neutron stars are slowly spinning, during the early part of the
evolution \citep{CF1994,LIGO}.  

Mass measurements from inspiralling binaries will be useful,
especially in constraining the equation of state through limits to the
neutron star maximum and minimum masses, but constraints to the radius
would be much more effective in constraining the nuclear equation of
state \citep{LP2001}.  Large finite-size effects, such as mass exchange
and tidal disruption, are measurable toward the end of inspiral
\citep{BC1992}, but the gravitational wave signal is expected to be
very complex during this period. \citet{FH2008} have recently pointed
out that tidal effects are also potentially measurable during the
early part of the evolution when the waveform is relatively clean.
The tidal fields induce quadrupole moments on the neutron
stars. This response of each star to external disturbance is described
by the Love number $k_2$ \citep{L1909}, which is a
dimensionless coefficient given by the ratio of the induced quadrupole
moment $Q_{ij}$ and the applied tidal field $E_{ij}$
\begin{equation}
	\label{k2}
	Q_{ij}=-k_2\frac{2 R^5}{3 G}E_{ij}\equiv-\lambda E_{ij} \, ,
\end{equation}
where $R$ is the radius of the star and $G$ is the gravitational
constant. The tidal Love number $k_2$, which is dimensionless, depends
on the structure of the star and therefore on the mass and the equation of
state (EOS) of dense matter.  The quantity $\lambda$ is the induced quadrupole
polarizability.

Tidal effects will form a very small correction in which the
accumulated phase shift can be characterized by a single quantity
$\bar\lambda$ which is a weighted average of the induced quadrupole
polarizabilities for the individual stars, $\lambda_1$ and
$\lambda_2$.  Since both neutron stars have the same equation of
state, the weighted average $\bar\lambda({\cal M})$, as a function of
chirp mass ${\cal M}=m_1^{3/5}m_2^{3/5}/(m_1+m_2)^{1/5}$, is
relatively insensitive to the mass ratio $m_1/m_2$, as is shown by
\citet{Hind2009}.  We therefore focus on the behavior of the
quadrupole polarizability $\lambda$ of individual stars.  These are
related to the dimensionless tidal Love number $k_2$ for each star by
$k_2=(3/2)G\lambda R^{-5}$.  The Love number $k_2$ is sensitive to the
neutron star equation of state, in particular to the compactness
parameter $M/R$ as shown by \citet{DN2009} and the overall
compressibility of the equation of state.  In particular, the tidal
Love numbers of strange quark matter stars are qualitatively different
from those of normal matter stars.  In a fashion similar to moment of
inertia measurements from relativistic binary pulsars \citep{LS2005},
an important constraint to the neutron star radius might become
possible from gravitational wave observations. Detection of the tidal
signature from coalescing compact binaries might provide an important,
and possibly unique, way to distinguish self-bound strange quark
matter stars from normal neutron stars.

Our paper is organized as follows.  In Sec. I, a new technique for the computation of tidal Love numbers is described. The influence of density discontinuities and phase transitions on Love numbers is discussed in Sec. II. Results of Love numbers for polytropic equations of state are presented in Sec. IV.   
Sec. V contains results for select analytic solutions of Einstein's equations  in spherical symmetry.  Love numbers for proposed model equations of state for normal stars with hadronic matter and self-bound stars with strange quark matter with and without crusts are given in Sec. VI,  wherein a comparison of results between these two distinct classes of stars are also made. In Sec VII, we discuss the role of a solid crust on Love numbers. Our results and conclusions are summarized in Sec. VII. Relevant parameters required for the computation of Love numbers for analytic solutions of Einstein's equations (discussed in Sec. V) are  to be found in Appendix A.

\section{Computation of Tidal Love Numbers}

The computation of tidal Love numbers is described by \citet{TC1967},
\citet{H2008}, \citet{DN2009}.  We use units in which $G=c=1$.  In
terms of the dimensionless compactness parameter $\beta=M/R$, the Love
number is given by
\begin{eqnarray}\nonumber
k_2(\beta,y_R)&=&\frac{8}{5}\beta^5(1-2\beta)^2\left[2-y_R+2\beta(y_R-1)\right]
\times{}\\ 
&\times&\{2\beta\left(6-3y_R+3\beta(5y_R-8)+2\beta^2\left[13-11y_R
  +\beta(3y_R-2)+2\beta^2(1+y_R)\right]\right) \\ 
\nonumber&+&3(1-2\beta)^2\left[2-y_R+2\beta(y_R-1)\}\log(1-2\beta)\right]^{-1}.
	\label{k2C}
\end{eqnarray}
Here, $y_R=[rH^\prime(r)/H(r)]_{r=R}$, where the function $H(r)$ is
the solution of the differential equation
\begin{equation}\label{eqH}\nonumber
H^{\prime\prime}(r)+H^\prime(r)
\left[\frac{2}{r}+e^{\lambda(r)}\left(\frac{2 m(r)}{r^2}+4\pi r(p(r)-
\rho(r))\right)\right]+H(r)Q(r)=0\,,
\end{equation}
where the primes denote derivatives with respect to $r$, and
\begin{equation}
Q(r)=4\pi e^{\lambda(r)} 
\left(5\rho(r)+9p(r)+\frac{\rho(r)+p(r)}{c_s^2(r)}\right)-
6\frac{e^{\lambda(r)}}{r^2}-\left(\nu^\prime(r)\right)^2\,.
\label{eq:Q}\end{equation}  
The metric functions $\lambda(r)$ and $\nu(r)$ for the spherical star are
\begin{equation}
e^{\lambda(r)}=\left[1-{2m(r)\over r}\right]^{-1}\,,\qquad
\nu^\prime(r)=2e^{\lambda(r)}{m(r)+4\pi p(r)r^3\over r^2}\,,
\label{eq:met}
\end{equation}
and $c_s^2(r)\equiv dp/d\rho$ is the squared sound speed.
Care has to be taken in the event of a first order phase transition or a 
surface density discontinuity in the evaluation of Eq. (\ref{eqH}) because 
the speed of sound vanishes.  We address this situation in the next section.

  We note that the calculation of the tidal Love number is simplified
by casting Eq. (\ref{eqH}) as a first-order differential equation for
$y(r)=rH^\prime(r)/H(r)$:
\begin{equation}
\label{eq:y}
ry^\prime(r)+y(r)^2+y(r)e^{\lambda(r)}\left[1+4\pi r^2(p(r)-\rho(r))\right]+
r^2Q(r)=0\,,
\end{equation}
so that it is necessary only to determine $y_R\equiv y(R)$; the value
of $H(R)$ is irrelevant.  The boundary condition for Eq. (\ref{eq:y})
is $y(0)=2$.  

\citet{DN2009} have emphasized that the factor $(1-2\beta)^2$ multiplying 
Eq. (\ref{k2C}) makes $k_2$ decrease rapidly with compactness $\beta$.  
Additionally, we note that for small compactness parameter $\beta$, there
are severe cancellations in Eq. (\ref{k2C}), and it is useful to expand
it in a Taylor series for $\beta<0.1$:
\begin{eqnarray}
k_2(\beta,y_R)=&{(1-2\beta)^2\over2}\Biggl[{2-y_R\over3+y_R}+
{y_R^2-6y_R-6\over(y_R+3)^2}\beta+{y_R^3+34y_R^2-8y_R+12\over7(y_R+3)^3}\beta^2
+{y_R^4+62y_R^3+84y_R^2+48y_R+36\over7(y_R+3)^4}\beta^4
\\ \nonumber
&+{5\over294}{5y_R^5+490y_R^4+1272y_R^3+1884y_R^2+1476y_R + 648\over(y_R+3)^5}\beta^5+\cdots\Biggr]
\label{eq:k2approx}
\end{eqnarray}
%

Note that in the Newtonian limit, $\beta\rightarrow0$, we have $p<<\rho, \rho
r^2<<1$, and one finds
\begin{eqnarray}\nonumber
ry^\prime(r)&+&y(r)^2+y(r)-6+4\pi r^2{\rho(r)\over c_s^2(r)}=0\,,\\
k_2(y_R)&=&{1\over2}\left({2-y_R\over3+y_R}\right)\,.
\label{eq:Newt}\end{eqnarray}

Equation (\ref{eq:y}) for $y$ must be integrated with the 
relativistic stellar structure, or TOV, equations: 
\citep{To1939,OV1939}
\begin{equation}
{dp(r)\over dr}=-{\left[m(r)+4\pi r^3p(r)\right]\left[\rho(r)+p(r)\right]\over
r(r-2m(r))}\,,\qquad{dm(r)\over dr}=4\pi\rho(r)r^2\,.
\label{eq:tov}
\end{equation}
We find it convenient to employ a thermodynamic variable $h(r)$, defined by
\begin{equation}
dh(r)={dp(r)\over\rho(r)+p(r)}\,,
\end{equation}
as the independent variable in place of $r$.  A stellar model can be
computed specifying the value of $h(0)$ at the star's center and
integrating equations for $dr/dh$ and $dm/dh$.  However, since these
equations are divergent at the origin and at the stellar surface, we
employed the radial variable $z=r^2$ instead.  One therefore has
\begin{eqnarray}
{dz\over dh}&=&-2{z(\sqrt{z}-2m)\over m+4\pi p z^{3/2}}\,,\cr
{dm\over dh}&=&2\pi\rho\sqrt{z}{dz\over dh}\,,\cr
\frac{dy}{dh} \nu'(h) \sqrt{z(h)}/2&=& y^2+y e^{\lambda(h)} \left(1+4 \pi z(h)(p(h)-\rho(h))\right)+z(h) Q(h)\,,
\label{eq:tov1}
\end{eqnarray}
where $Q$ is determined by Eq. (\ref{eq:Q}).  
The behavior of $y$ near the star's center is given by
\begin{equation}
	\label{yin}
	y(h)=2-\frac{6}{7}\frac{5 \rho_c+9 p_c+(p_c+\rho_c)/c_{sc}^2}{3 p_c+\rho_c} (h_c-h)+O\left((h_c-h)^2\right).
\end{equation}
Also note that $y_R\equiv y(h=0)$.

In some cases, such as with polytropic equations of state, we found it
was better to use $\ln h$ as the independent variable.  In addition,
some care has to be taken in the event that $d\rho/dh$ diverges at the
stellar surface, which is the case for polytropes if the polytropic
index $n<1$.

\section{The role of density discontinuities and phase transitions}

As Eq. (\ref{eq:y}) for $y$ contains the squared adiabatic speed of
sound $c_s^2=dp/d\rho$, the solution will be altered in the case of
phase transitions within the star, for example, between the crust and
the core, or in the case of a finite surface density such as appears
in models of strange quark stars or for a uniform density stellar
model.  However, in the event that multiple charges (e.g., electric
charge and baryon number) are conserved in a phase transition, the
constraint of global charge neutrality (two Gibb's phase rules)
results in a continuous pressure versus energy density curve even if
the phase transition is of first order.  The situation of a density
discontinuity was elaborated in by \citet{DN2009}, who showed that a large
discontinuity in
the energy density will greatly change the value of $k_2$.

Expressing the sound speed in the vicinity of a density
discontinuity as
\begin{equation}
	\label{eqdendis}
	\frac{d\rho}{dp}=\frac{1}{c_s^2}=\left. \frac{d\rho}{dp}
	\right|_{p \neq p_d}+\Delta \rho \, \delta(p-p_d),
	\end{equation} 
where $p_d$ is the pressure at the discontinuity and $\Delta
\rho=\rho(p_d+0)-\rho(p_d-0)$ is the energy density jump across the
discontinuity.  While solving Eqs. (\ref{eq:tov1}), this discontinuity
can be taken into account by properly matching solutions at the point
of discontinuity $r_d=r(h_d)$:
\begin{equation}
	y(r_d+\epsilon)=y(r_d-\epsilon)-\frac{\rho(r_d+\epsilon)-\rho(r_d-\epsilon)}{m(r_d)/(4 \pi r_d^3)}=y(h_d-\epsilon)-3\frac{\Delta \rho}{\tilde{\rho}},
	\label{dis_r_jump}
\end{equation}
where $\epsilon \to 0$ and $\tilde{\rho} = m(r_d)/(4 \pi r_d^3/3)$ is
the average energy density of the inner ($r<r_d$) core.

\section{Polytropic Equations of State}

\begin{figure}[h]
	\includegraphics[width=8cm,angle=90]{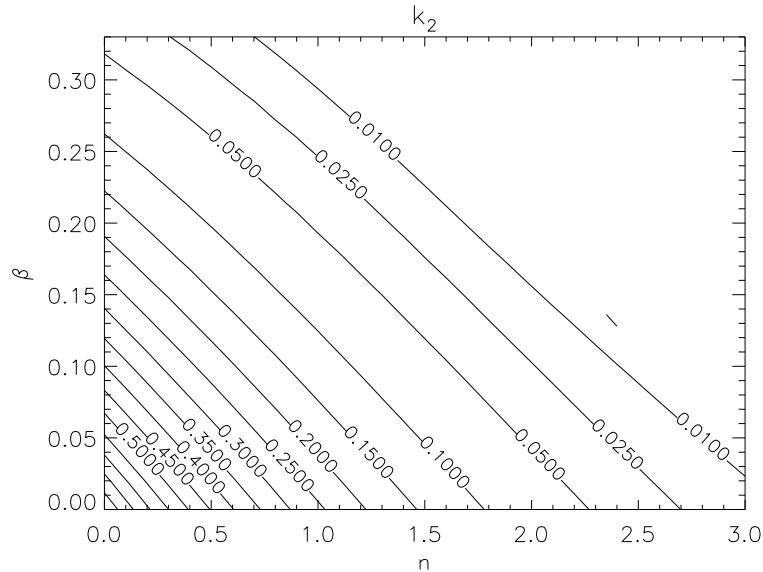}
	\caption{Contours of the dimensionless tidal Love number $k_2$ as a
          function of compactness $\beta=M/R$ and polytropic index $n$ 
	(labelled along curves) 
          for polytropes.  Contours are not shown for configurations
          that are hydrostatically unstable (i.e., those with central densities
          larger than that of the maximum mass).}
	\label{fig:k2poly}
\end{figure}
\begin{figure}[h]
	\includegraphics[width=8cm,angle=90]{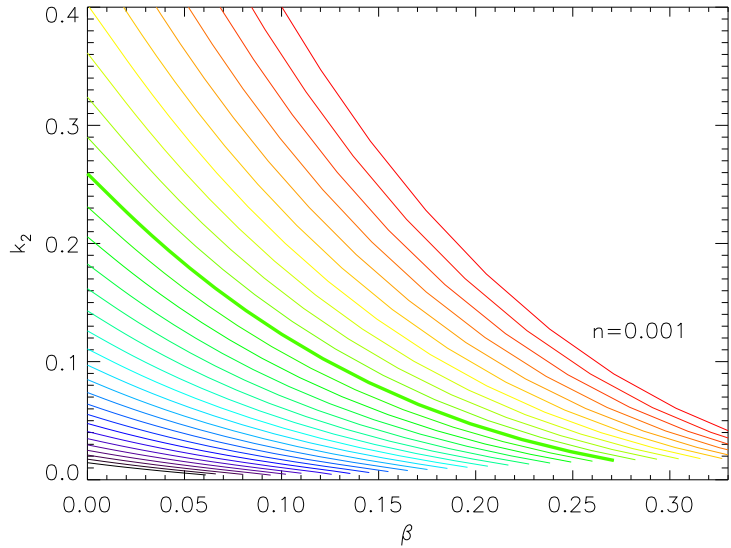}
	\caption{The dimensionless tidal Love number $k_2$ as a
          function of compactness $\beta=M/R$ and polytropic index $n$
          for polytropes.  The polytropic index $n=0.001$ for the
          top-most curve and in multiples of 0.1 for each succeeding
          curve.  The thickest curve shows results for $n=1$.}
	\label{fig:k2p}
\end{figure}
\begin{figure}[h]
	\includegraphics[width=8cm,angle=90]{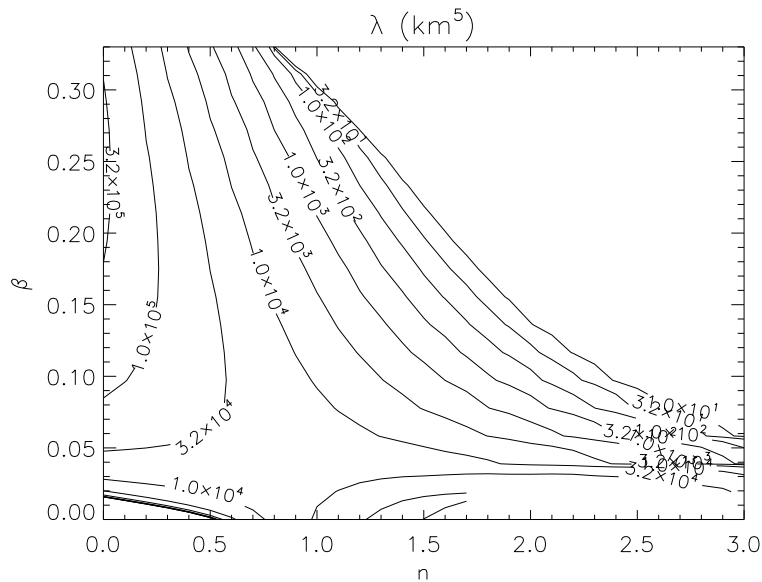}
	\caption{The quantity $\lambda=(2G/3)k_2 R^5$, in units of
          km$^5$, as a function of compactness $\beta=M/R$ for
          polytropes of index $n$.  
	  Contours are not shown for configurations that
          are hydrostatically unstable.}
	\label{fig:k2R5poly}
\end{figure}
\begin{figure}[h]
	\includegraphics[width=8cm,angle=90]{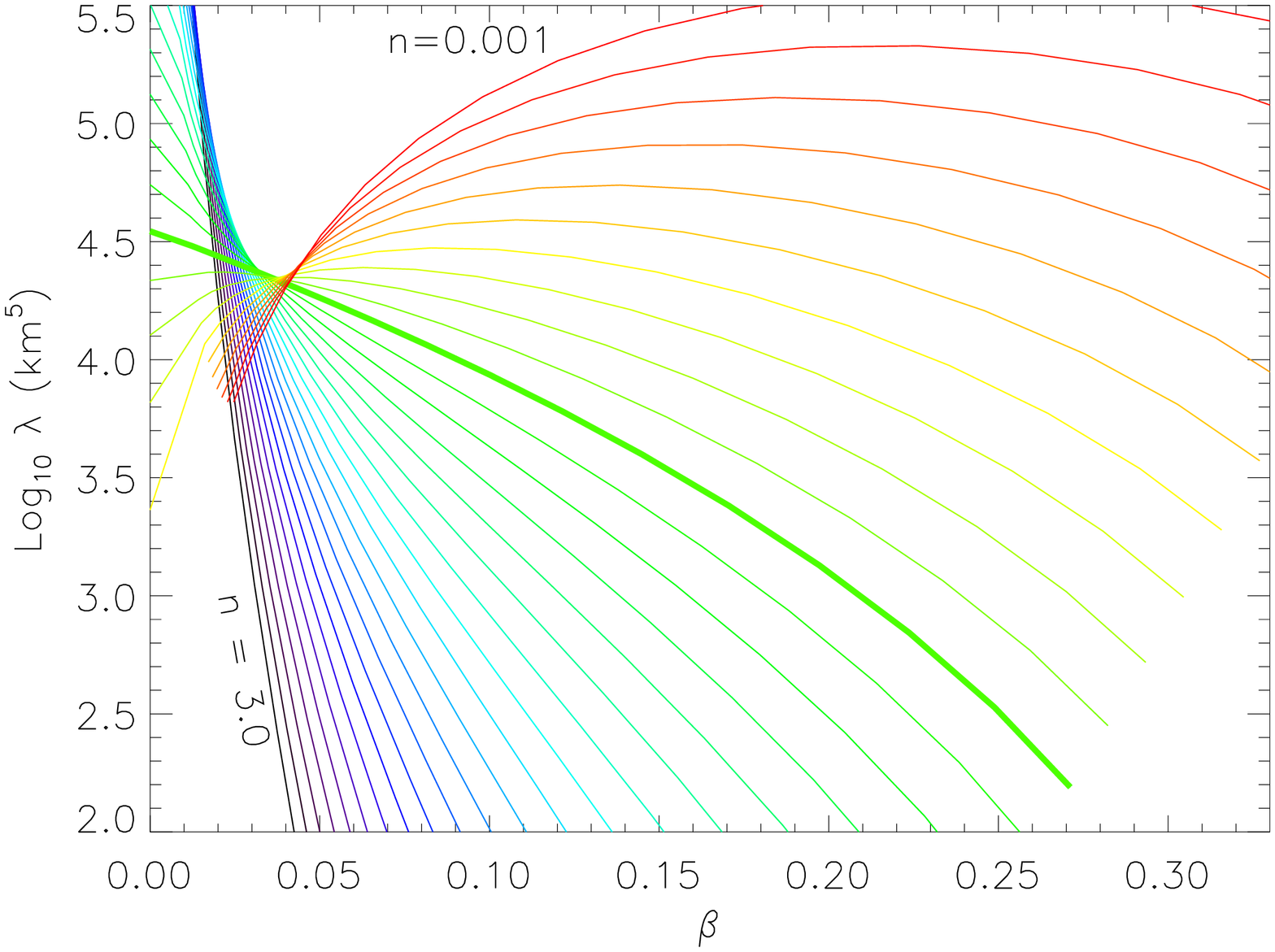}
	\caption{The quantity $\lambda=(2G/3)k_2 R^5$, in units of
          km$^5$, as a function of compactness $\beta=M/R$ for
          polytropes ranging from $n=0.001$ (top-most curve) to 3.0 
	  (left-most curve) in increments of 0.1. Results for the
          polytrope $n=1$ are shown as a thick curve.}
	\label{fig:lambdapoly}
\end{figure}
It is useful to evaluate tidal Love numbers for polytropic equations
of state $p=K\rho^{1+1/n}$.  Love numbers in the Newtonian limit for
polytropes have been calculated by \citet{BO1955} and \citet{KS1995}.
In the Newtonian limit, it is easily observed that the values for $y$
and $k_2$ are independent of the polytropic constant
$K=p/\rho^{1+1/n}$, which scales out of Eq. (\ref{eq:Newt}).  However,
the quadrupole polarizability $\lambda=(2/3)k_2R^5$, and therefore the
gravitational wave signature, does depend on $K$.  There exist analytic
solutions for the Newtonian case for polytropes of indices $n=0$ and
1.  In the case $n=0$, an incompressible fluid, $c_s^2=\infty$ and the
solution inside the star which satisfies the boundary condition at the
center is simply $y(r)=2$.  However, the discontinuity in the sound
speed at the stellar surface must be taken into account.  According to
Eq. (\ref{dis_r_jump}), $y_R$ receives a boundary contribution $4\pi
R^3\rho/M=3$, where $\rho$ is the constant energy density inside the
star.  Therefore, for an incompressible fluid, $y_R=y(r_-)-3=-1$ and
$k_2=3/4$.

In the case $n=1$, one finds \citep{H2008}
\begin{equation}
y(r)={\pi r\over R}{J_{3/2}(\pi r/R)\over J_{5/2}(\pi r/R)}-3\,,
\qquad y_R={\pi^2-9\over3}\,,\qquad k_2={15-\pi^2\over2\pi^2}\,,
\hspace*{2cm}n=1.
\label{eq:n=1}
\end{equation}
In the above, $J_i(x)$ is the standard Bessel function. 

\citet{H2008, DN2009, BP2009} have examined relativistic polytropic
equations of state in the case of finite compactnesss.  We have repeated
these calculations.  For each $n$,
the polytropic constant $K$ was determined from the fiducial pressure
$p_0=1.322\times 10^{-6}$ km$^{-2}$ and $\rho_0=1.249\times 10^{-4}$
km$^{-2}$ using $K=p_o\rho_0^{-1-1/n}$. These values are equivalent
to the pressure $p_0=1$ MeV fm$^{-3}$ and mass-energy density
$\rho_0=94.38$ MeV fm$^{-3}$ (or a baryon density $n_0=0.1$ fm$^{-3}$
for the case $n=1$).  These values
were chosen to produce reasonable neutron star radii for solar mass
neutron stars.  For soft EOS's, $n>1$, the stellar radius
decreases with increasing mass up to the maximum mass and the maximum
mass stars are relatively lighter than for stiff EOS's, $n<1$.  For
$n<1$, the stellar radius generally increases with increasing mass
until the maximum mass is approached.  The case $n=1$ is intermediate
and has a finite radius even for a star with vanishing mass.

The results of integrating Eq. (\ref{eq:y}) for these polytropic EOS's
are summarized in Figs. \ref{fig:k2poly} and \ref{fig:k2p} which show
$k_2$ as a function of $\beta$ and $n$.  Generally, $k_2$ decreases
with increasing $n$ and $\beta$.  The gravitational response is
proportional to $\lambda=(2G/3)k_2R^5$ and this is shown for relativistic
polytropes in Figs. \ref{fig:k2R5poly} and \ref{fig:lambdapoly}.  This
quantity decreases rapidly with increasing $n$, and for $n\ge0.5$, it
also decreases rapidly with the compactness parameter $\beta$.

We have found that the results for $k_2$ do not significantly depend
on the value $K$ in the relativistic case by altering our fiducial values
of $p_o$ or $\rho_o$ within reasonable ranges resulting in configurations of
similar dimensions to neutron stars.  Our results are the same
as those of \citet{H2008, DN2009, BP2009} to within numerical
accuracy.

\section{Love Numbers for Analytic Solutions of Einstein's Equations}

\begin{figure}[h]
	\includegraphics[width=8cm,angle=90]{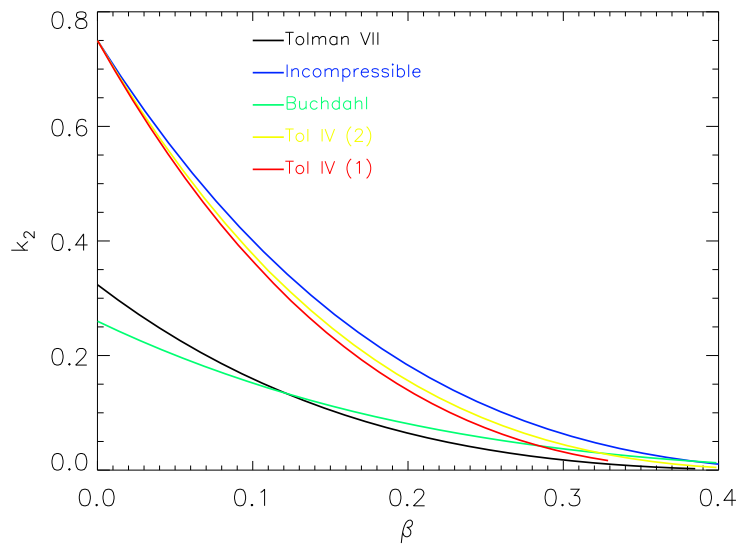}
	\caption{The dimensionless tidal Love number $k_2$ as a
          function of compactness $\beta=M/R$ for analytic solutions 
	(see Appendix A) of
Einstein's equations in spherical symmetry.}
	\label{fig:k2anal}
\end{figure}
It is also useful to compute the tidal response for some of the known
analytic solutions of Einstein's equations in spherical symmetry.  All
analytical solutions are scale-free; they contain essentially two
parameters, the central energy density $\rho_c$ and compactness
parameter $\beta=GM/Rc^2$.  Among the useful analytic solutions we
will study are (i) the uniform fluid sphere, (ii) the Tolman VII
solution~\citep{To1939}, (iii) Buchdahl's
solution~\citep{Bu1959,Bu1966}, and (iv) and (v), two generalizations of
the Tolman IV solution~\citep{Na1950,Na1951,La2007}.  The Tolman VII
and Buchdahl's solutions have vanishing surface energy densities and
are useful approximations to realistic neutron star models.  The
incompressible fluid and the generalizations of the Tolman IV solution
have finite surface densities, and the latter are reasonable
approximations of strange quark matter stars.

It is useful to recast Eq. (\ref{eq:tov1}) in
the form
\begin{eqnarray}
{dw\over dh}&=&-2{w(\sqrt{w}-2x\beta)\over x\beta+\alpha(p/\rho_c)w^{3/2}}\,,\qquad
{dx\over dh}={dw\over dh}\left[{\alpha\over2\beta}{\rho\over\rho_c}\sqrt{w}\right]\,,\cr
{dy\over dh}&=&{dw\over dh}\left\{-{y^2+ye^\lambda-6e^\lambda\over2w}+{\alpha\over2}e^\lambda
\left[\left({\rho\over\rho_c}-{p\over\rho_c}\right)y-5{\rho\over\rho_c}-
9{p\over\rho_c}-{\rho+p\over\rho_cc_s^2}\right]+
{2\over w}e^{2\lambda}\left({1-e^{-\lambda}\over2} + 
\alpha w{p\over\rho_c}\right)^2\right\}\,,
\label{eq:tov2}
\end{eqnarray}
where $\alpha=4\pi\rho_c R^2, x=m/M, \beta=M/R$ and $w=r^2/R^2$.
Therefore, we need the quantities $\rho/\rho_c$, $p/\rho_c$, $c_s^2$,
$\alpha$ and $e^\lambda$ for each analytic equation of state.  In
addition, for the Tolman IV solutions, which have a finite surface
density, the boundary contribution to $y_R$ is required.  This
quantity, in the present notation, is
$-(\alpha/\beta)(\rho_s/\rho_c)$.  The quantity $\rho_s/\rho_c$
together with the above quantities are provided in Appendix A.

As shown in Fig. \ref{fig:k2anal}, the two analytic solutions that most closely resemble normal neutron
stars, the Buchdahl and Tolman VII solutions, predict values of $k_2$
that are similar and which closely track the results for the $n=1$
polytrope (of course, for $\beta=0$, Buchdahl's solution and the $n=1$
polytrope are identical).  In contrast, the Incompressible and Tolman
IV solutions represent a significantly different family, and, as we
will see, are good approximations to strange quark matter stars.  It is
clear that the two families of analytic solutions have different
behaviors, and this foreshadows the results for the equation of state models we
discuss below.  Because of the scale-free character of these
solutions, we have not shown results for $\lambda$, which will scale
with the assumed $\rho_c$ (or, equivalently, $M$ or $R$.)

\section{Love Numbers for Model Equations of State}
\subsection{Hadronic Equations of State}

\begin{figure}[h]
	\includegraphics[width=9cm,angle=90]{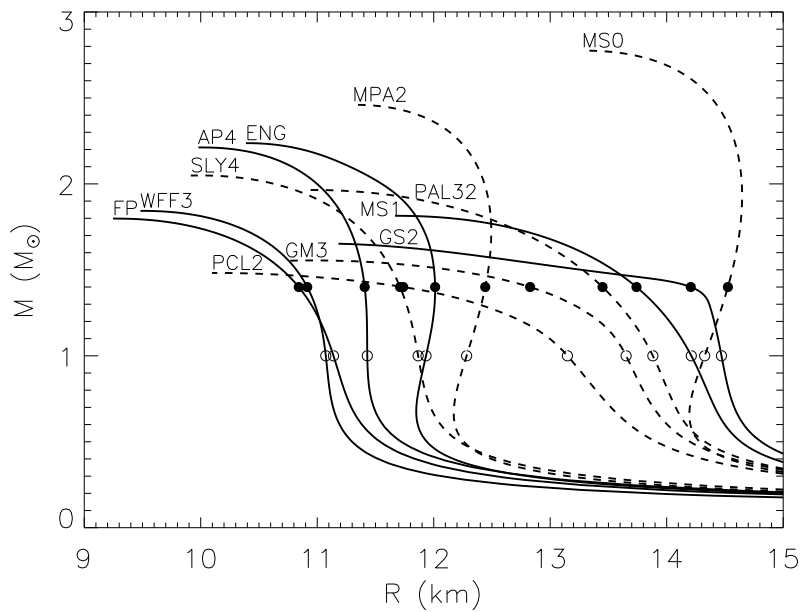}
	\caption{Mass-radius diagram for the hadronic equation of
          states used in this paper.  Filled (open) circles indicate
          configurations with $M=1.4$ M$_\odot$ (1.0 M$_\odot$). The
          EOS notation follows \citet{LP2001} and Table I.}
	\label{fig:hadronMR}
\end{figure}

\begin{table}
\centerline{EQUATIONS OF STATE}
\begin{tabular}{l|l|l|l} \hline\hline
Symbol & Reference & Approach & Comp. \\ \hline
FP  & Friedman \& Pandharipande  & Variational & np \\
WFF(1-3) & Wiringa, Fiks \& Fabrocine & Variational & np \\
AP(1-4) & Akmal \& Pandharipande & Variational & np \\
MS(0-3) & M\"uller \& Serot & Field Theoretical & np \\   
MPA(1-2) & Mu\"ther, Prakash \& Ainsworth & Dirac-Brueckner HF & np \\
ENG & Engvik et al. & Dirac-Brueckner HF & np \\
PAL(1-6)  & Prakash, Ainsworth \& Lattimer & Schematic Potential & np \\
GM(1-3) & Glendenning \& Moszkowski & Field Theoretical & npH \\
GS(1-2) & Glendenning \& Schaffner-Bielich & Field Theoretical & npK\\
PCL(1-2) & Prakash, Cooke \& Lattimer~\citep{sqm123} & Field Theoretical & npHQ\\ 
SLY4 & Douchin \& Haensel~\citep{D2001}& Field Theoretical & npe\\ 
SQM(1-3) & Prakash, Cooke \& Lattimer~\citep{sqm123} & Quark Matter & Q $(u,d,s)$\\
STE & Steiner, Fig. \ref{EOSsqm} & Quark Matter & Q $(u,d,s)$\\
PAG & Page, Fig. \ref{EOSsqm} & Quark Matter & Q $(u,d,s)$\\
ALF & Alford, Fig. \ref{EOSsqm} & Quark Matter & Q $(u,d,s)$\\
\hline
HS & Haensel, Salgado \& Bonazzola~\citep{Haensel95} & Crust & Z,e,n\\ 
BPS & Baym, Pethick \& Sutherland~\citep{Baym1971} & Crust & Z,e,n\\
\hline
\hline
\end{tabular}
\label{eosnames}
\caption{Approach refers to the
underlying theoretical technique.  Composition (Comp.) refers to strongly
interacting components (n=neutron, p=proton, Z=nucleus, H=hyperon, K=kaon,
Q=quark); all models include leptonic contributions. This table is slightly expanded from the version found in \citep{LattimerPrakash2001} which contains references not noted here.}
\end{table}
The hadronic EOS's were taken from a compilation by \citet{LP2001}
 that describes their origins.  There are three generic families of
 equations of state: (i) normal nucleonic equations of state, (ii)
 equations of state with considerable softening above the nuclear
 saturation density, due to Bose condensation, hyperons or a mixed
 quark-hadronic phase, and iii) strange quark matter stars. We have
 used a selection in an attempt to span the extreme range of models of
 each type.
The mass-radius curves for hadronic EOS's are shown in
Fig. \ref{fig:hadronMR}. 
\begin{figure}[h]
	\includegraphics[width=8.8cm,angle=90]{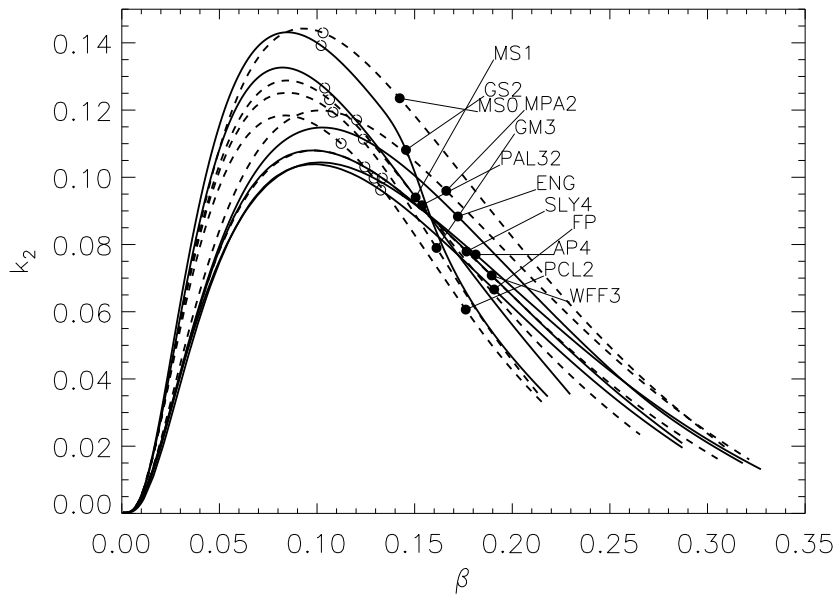}
	\caption{The dimensionless tidal Love number $k_2$ as a
          function of compactness $\beta=M/R$ for hadronic EOSs.   
	Filled (open) circles indicate
          configurations with $M=1.4$ M$_\odot$ (1.0 M$_\odot$). The
          EOS notation follows \citet{LP2001} and Table I.}
	\label{fig:hadronk2beta}
\end{figure}
\begin{figure}[h]
	\includegraphics[width=8.8cm,angle=90]{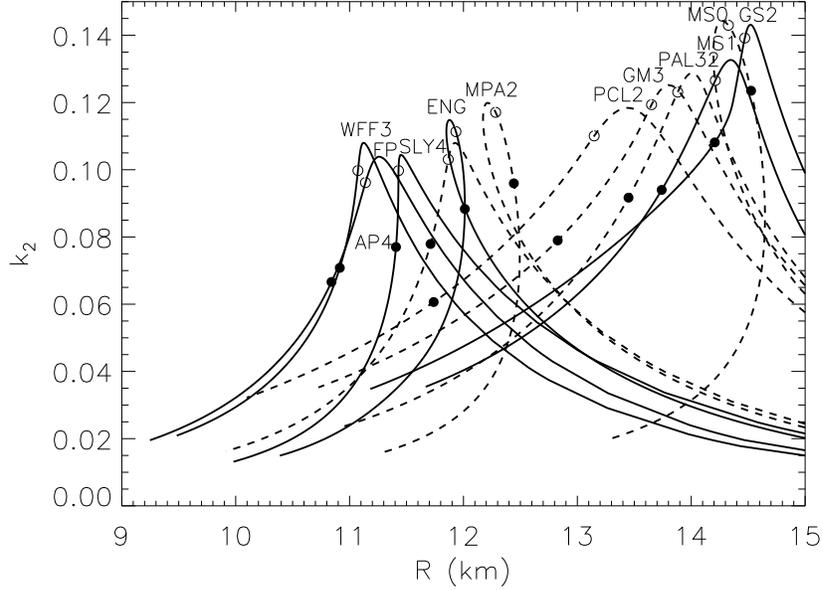}
	\caption{The Love number $k_2$ as a function of radius $R$.   Filled (open) circles indicate
          configurations with $M=1.4$ M$_\odot$ (1.0 M$_\odot$). The
          EOS notation follows \citet{LP2001} and Table I.}
	\label{hadronk2R}
\end{figure}
\begin{figure}[h]
	\includegraphics[width=8.8cm,angle=90]{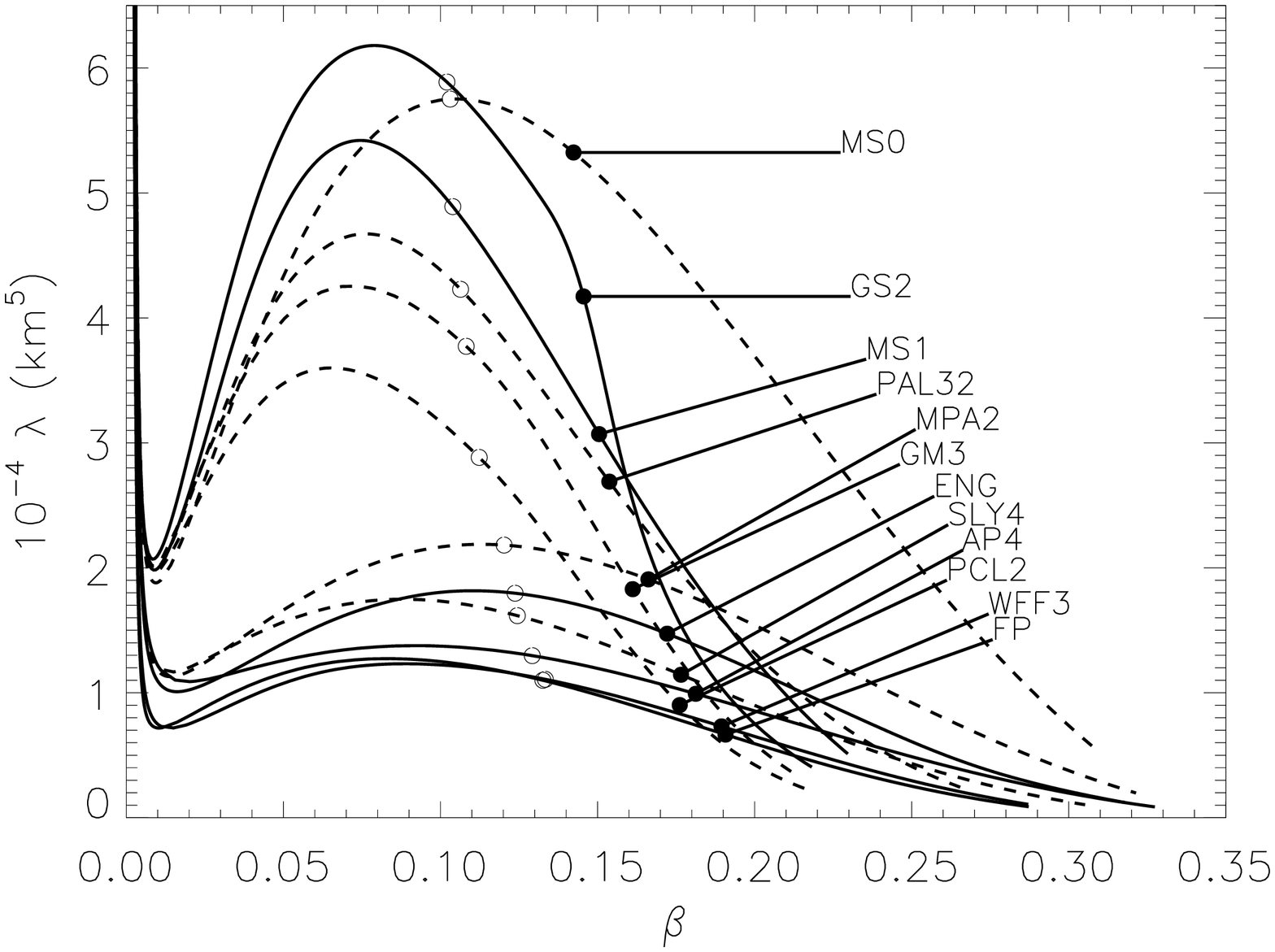}
	\caption{The quantity $\lambda=(2/3)k_2R^5$ for hadronic
          equations of state.    Filled (open) circles indicate
          configurations with $M=1.4$ M$_\odot$ (1.0 M$_\odot$). }
	\label{fig:hadronlambdabeta}
\end{figure}
\begin{figure}[h]
	\includegraphics[width=8.8cm,angle=90]{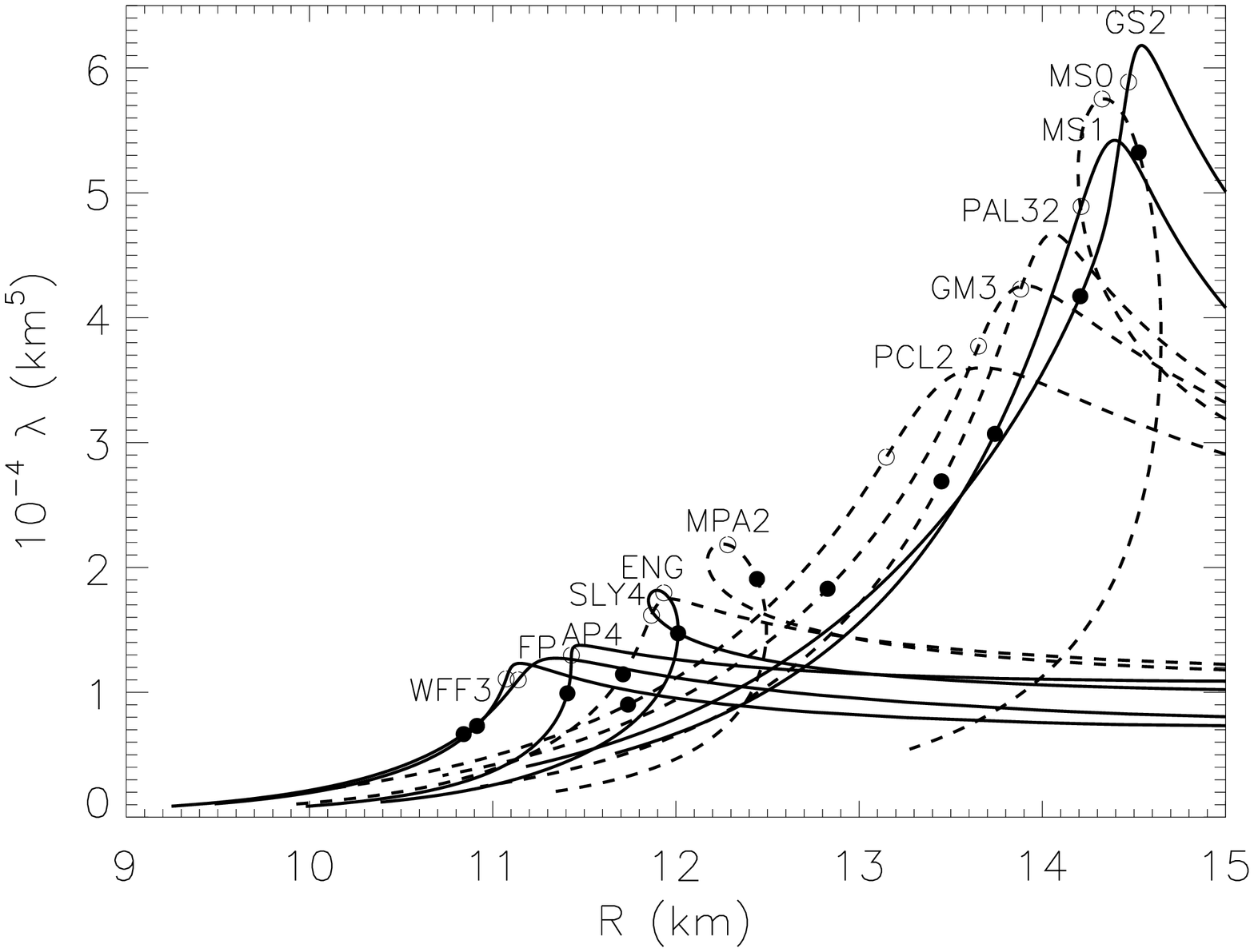}
	\caption{The quantity $\lambda=(2/3)k_2R^5$ for hadronic
          equations of state.    Filled (open) circles indicate
          configurations with $M=1.4$ M$_\odot$ (1.0 M$_\odot$). }
	\label{fig:hadronlambdaR}
\end{figure}

Love numbers as a function of compactness are shown in
Fig. \ref{fig:hadronk2beta} for hadronic models.  There is a
relatively narrow spread of values of $k_2$ for a given compactness,
and for each EOS, the value of $k_2$ appears to be a maximum for
masses near 1 M$_\odot$.  In contrast to the analytic Tolman VII and
Buchdahl solutions, for which $k_2(\beta\rightarrow0)\simeq0.3$, $k_2$
tends to zero for small $\beta$ for realistic equations of state.  The
fact that hadronic equations of state have a small range of variations
as a function of compactness is reminiscent of the situation for the
moment of inertia \citep{LS2005}.  

It is useful to examine $k_2$ as a function of neutron star radius, as
shown in Fig. \ref{hadronk2R}.  Although the range of values
observed for $k_2$ are common to all models, it is now clear that the
quadrupole response will vary more widely, due to it being
proportional to $R^5$.  In Figs. \ref{fig:hadronlambdabeta} and
\ref{fig:hadronlambdaR} the quadrupole response is shown.  The maxima in
$\lambda$ occurs near 1 M$_\odot$, as it did for $k_2$, and their is a
pronounced trend for $\lambda$ to increase with $R$.
Assuming the true neutron star equation of state is hadronic, it
therefore appears that a measurement of $\lambda$ translates into an
estimate of $R$ relatively independently of the details of the
equation of state.  In fact, compared to the moment of inertia which
scales as $R^2$, the potential for a radius constraint is enhanced due
the $R^5$ behavior of $\lambda$. 
\subsection{Self-bound strange quark matter stars}

\begin{figure}[tb]
	\includegraphics[width=12cm]{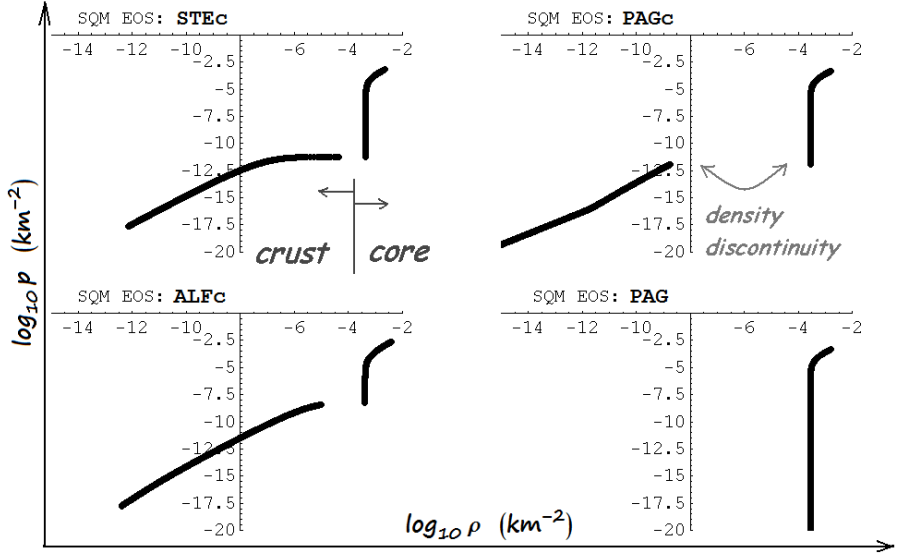}
	\caption{Pressure versus energy density for strange quark matter equations of state with and without crust. Equation of state STE is taken from Steiner~\citep{Andrew}, PAG from Page~\citep{Dany} and ALF from Alford~\citep{Mark} (see Table I). Density discontinuities are as indicated.}
\label{EOSsqm}     
\end{figure}
\begin{figure}[tb]
  \includegraphics[width=8.8cm,angle=90]{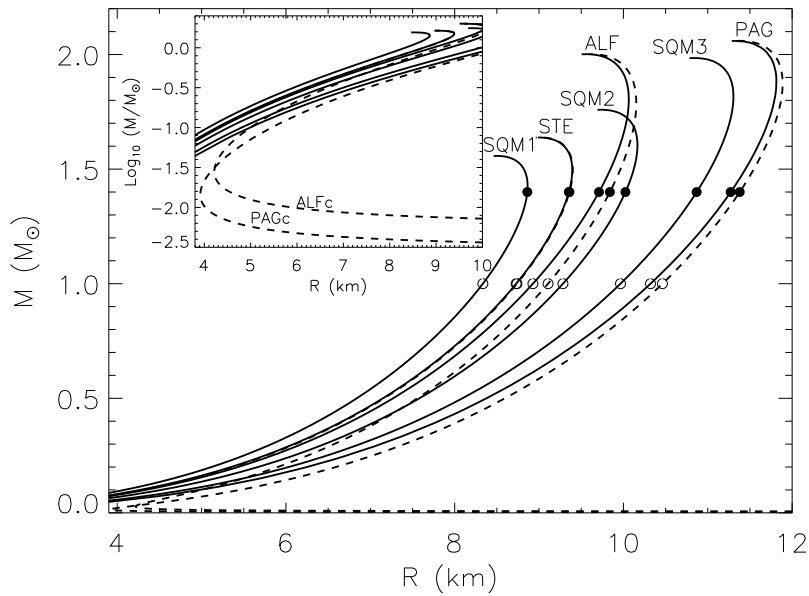}
	\caption{Mass-radius curves for strange quark matter
          equations of state. The insert shows
          results on a logarithmic scale to highlight the effects of a
          hadronic crust.}
          \label{quarkMR}     
\end{figure}
We turn now to examine results of Love numbers for self-bound strange
quark matter stars.  It is uncertain whether or not strange quark
matter stars will have significant crusts or not, so we examine models
of both kinds.  Models without crusts are characterized by quark
matter extending up to a bare surface with a finite baryon density of
2 to 3 times nuclear matter equilibrium density. Crusts of normal
matter on top of such stars might be supported by strong electric
fields at the surface.    Fig. \ref{EOSsqm} shows three examples for both cases
(STE from Steiner~\citep{Andrew}, PAG from
Page~\citep{Dany} and ALF from Alford~\citep{Mark}). The
crust and the core regions are apparent from the large discontinuity
in the energy density. The existence of a crust results in large radii
for small stellar masses (of order 0.01 M$_\odot$), but do not
dramatically affect the radii of stars with masses larger than 0.1
M$_\odot$ (see Fig. \ref{quarkMR}).  It therefore appears unlikely that
the existence of a crust has a pronounced effect on the Love number or
quadrupole properties of the star.

\begin{figure}[tb]
	\includegraphics[width=8.8cm,angle=90]{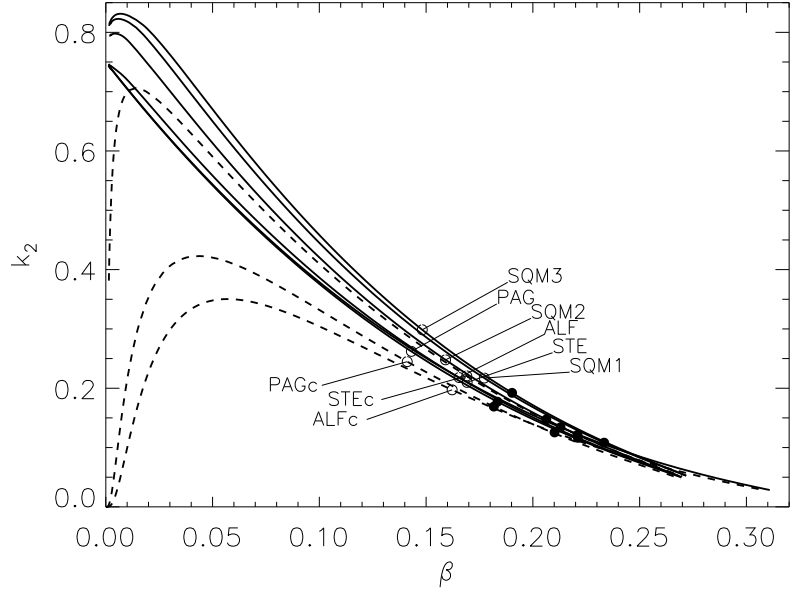}
	\caption{Dimensionless Love numbers for the strange quark
          matter stars. Filled (open) circles
          indicate configurations with $M=1.4$ M$_\odot$ (1.0
          M$_\odot$). }
          \label{quarkk2beta}     
\end{figure}
\begin{figure}[tb]
	\includegraphics[width=8.8cm,angle=90]{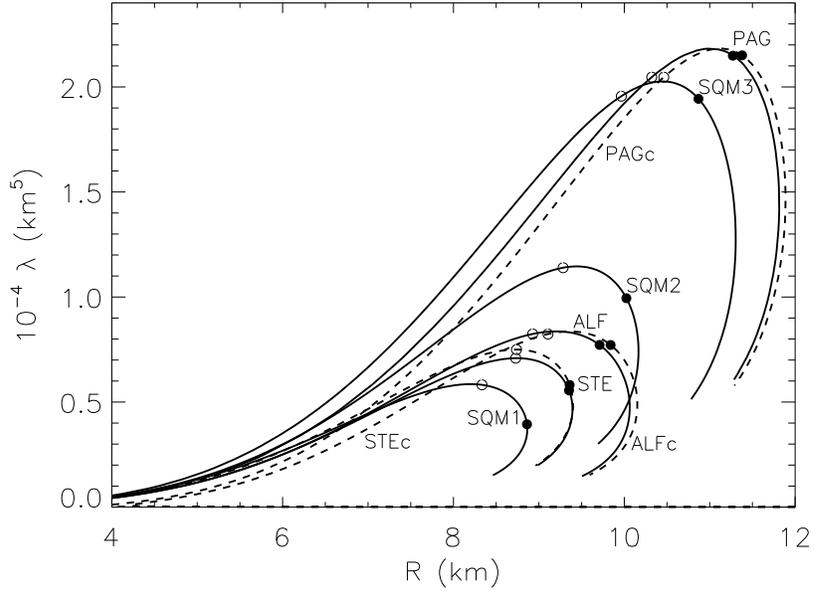}
	\caption{Quadrupole polarizabilities $\lambda$ for the strange quark
          matter stars. Filled (open) circles
          indicate configurations with $M=1.4$ M$_\odot$ (1.0
          M$_\odot$). }
          \label{quarklambdaR}     
\end{figure}

In Fig. \ref{quarkk2beta}, the dimensionless Love number $k_2$ is
shown as a function of compactness. As was the case for hadronic
stars, there is a clustering of curves relatively independent of the EOS
for stars without crusts.  The curves follow the analytic results for
the incompressible fluid and for the Tolman IV solutions, and differ
from hadronic cases by having a large, finite value of $k_2$ for small
$\beta$.  However, in the case of an added crust, $k_2$ is reduced at
small values of $M/R$, but this only occurs for ultra-low mass stars.   
For masses in excess of 1 M$_\odot$, the Love number approaches the 
corresonding values for  hadronic stars, and the effect of the crust is
negligible.

The quadrupole response $\lambda=2 k_2 R^5/3$ is shown in
Fig. \ref{quarklambdaR} as a function of radius. The strong dependence
on radius follows the trend noted for hadronic stars.  The effect of
the crust is unimportant.
\subsection{Comparison of normal and self-bound stars}

\begin{figure}[h]
	\includegraphics[width=8.8cm,angle=90]{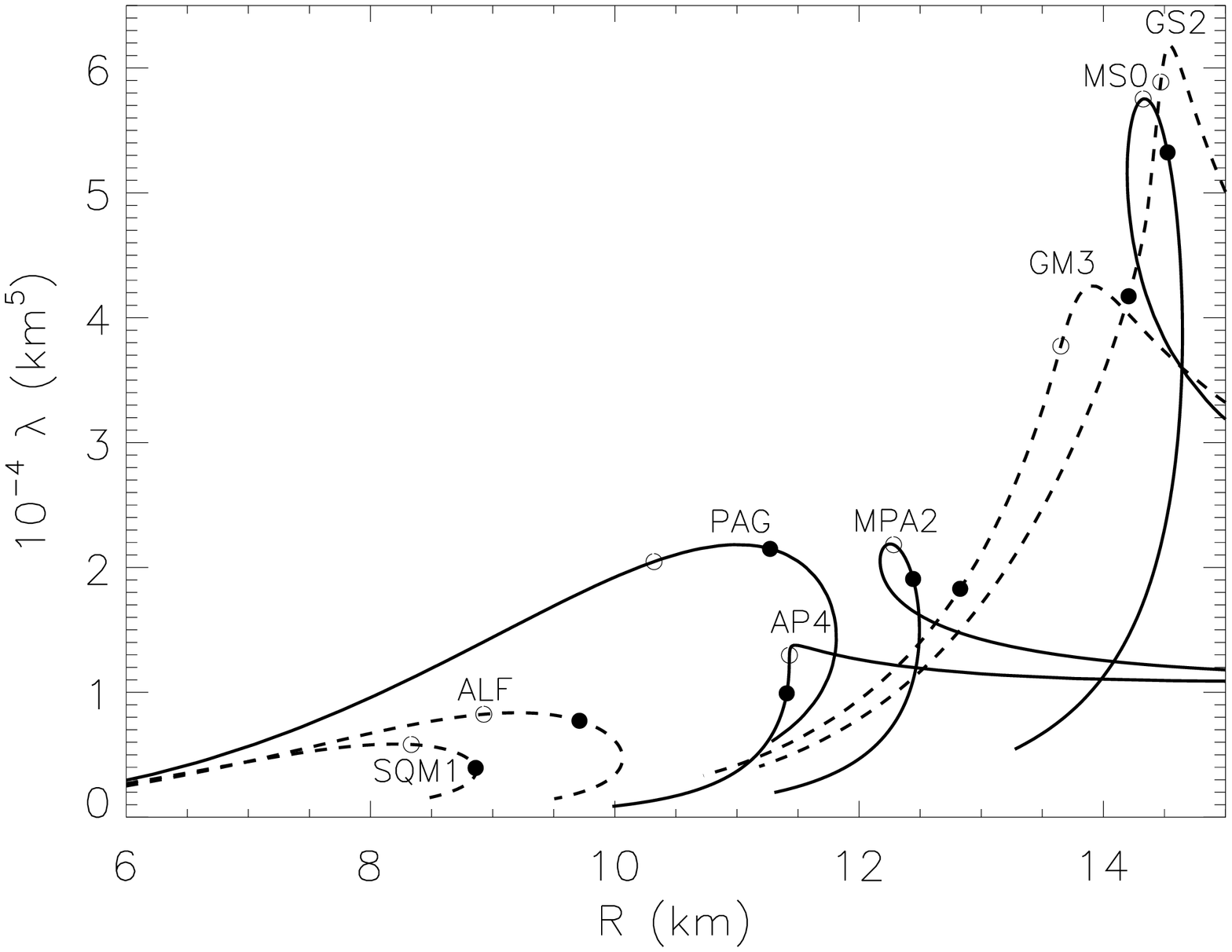}
	\caption{Comparison of quadrupole polarizabilities $\lambda$
          for normal and strange quark matter stars. Filled (open)
          circles indicate configurations with $M=1.4$ M$_\odot$ (1.0
          M$_\odot$). }
          \label{mixedlambdaR}     
\end{figure}
In order to elaborate the distinction between strange quark matter and
hadronic models, we show the quadrupole response $\lambda=2 k_2 R^5/3$
in Fig. \ref{mixedlambdaR} for a representative sample of models of
each type.  The strong dependence of $\lambda$ on $R$ is common to all
models.  Where the radii of models overlap, however, it appears that the
strange quark matter configurations have values of $\lambda$ about
50\% larger.  This difference is probably too small to be observable,
and it appears doubtful that any quark matter configurations will have
a strong enought tidal signature to be observed.

\section{Discussion}

The combined tidal effects of two neutron stars in circular orbit can be
found from a weighted average of the quadrupole responses \citep{FH2008}:
\begin{equation}
	\label{eq:lamt}
	\tilde{\lambda}=\frac{1}{26}\left((11 m_2+M)\frac{\lambda_1}{m_1}+(11 m_1+M)\frac{\lambda_2}{m_2}\right),
\end{equation}
where $M=m_1+m_2$ is the total mass of the binary and $\lambda_1$ and
$\lambda_2$ are the quadrupole responses of $m_1$ and $m_2$.  Note
that if $m_1=m_2$, then $\lambda_1=\lambda_2=\tilde{\lambda}$.  If
$m_2 =0.5 m_1$, then $\tilde{\lambda}\approx (40/26)\lambda_1$. It is
unlikely that the mass ratio would be smaller than this amount, as
the minimum neutron star mass that can be formed in supernovae is not
less than 1 M$_\odot$ and the maximum neutron star mass is of order 2
M$_\odot$.  Therefore, the value of $\bar\lambda$ is similar to that
of the largest neutron star.  In the case that the individual masses
can be found to reasonable accuracy from the gravitational wave
signal, the individual values of $\lambda$ for the two stars will be
determined to an accuracy constrained by the errors in $\bar\lambda$
and the masses.  

We have assumed in evaluating the Love numbers that the crust behaves
as a liquid.  However, if the stress on the solid crust produced by the
tidal field is large enough, then the crust can be melted and our
calculations become valid. The strength required to melt the crust
can be estimated from the results of recent work on crust
breaking. We estimate the induced quadrupole moment to be
\begin{equation}
  Q_{22}=\lambda E_{22}\approx\lambda\sqrt{E_{ij}E^{ij}}=\sqrt{3\over2}\lambda\frac{M}{D^3},
	\label{q22}
\end{equation}
where the tidal field strength $E_{ij}$ \citep{Favata2006} depends on 
the distance $D$ between the stars and $M$ is the total mass; we assumed
for simplicity an equal-mass binary. 
Assuming a binary in circular orbit, we can
calculate the orbital frequency $\Omega$ from Kepler's third law
\begin{equation}
  \Omega^2 \approx \frac{M}{D^3}. 
  \label{Omega}
\end{equation}
Eliminating $M/D^3$ using Eq. (\ref{q22}), and recognizing that the
frequency of the emitted gravitational waves $f$ is twice the orbital
frequency \citep{twice}, we have
\begin{equation}
  f= \frac{2}{2 \pi} \Omega \approx \frac{1}{\pi} \sqrt{\frac{Q_{22}}{\lambda}}\left(\frac{2}{3}\right)^{1/4},
  \label{gwf}
  \end{equation}
which has an implicit mass dependence through $Q_{22}$ and $\lambda$.
For a 1 M$_\odot$ neutron star using the EOS labelled SLY,
\citet{Horowitz2009b} estimates that the maximum value of $Q_{22}$
reached at the breaking point of the crust, where the strain
$\sigma\approx0.1$, \citet{Horowitz2009a}, is approximately
$Q_{22,max}=10^{40}$ g cm$^2$.  The breaking point is therefore reached
during the inspiral of an equal-mass binary at the moment when the
frequency of detected gravitational waves becomes
\begin{equation}
\label{fbrn}
f_{br} \approx \frac{(2/3)^{1/4}}{\pi} \left(\frac{10^{40} {\rm{\,g\,cm^2}}}{2 \, 10^{36} {\rm{\,g\,cm^2\,s^2}}}\right)^{1/2} \approx 20 {\rm{\,Hz}},
\end{equation}
where we used the value for $\lambda$ for a 1 M$_\odot$ star as
determined in Fig. \ref{fig:hadronlambdabeta}.  Note that this frequency
implies a binary separation distance $D_{br}\approx 400$ km from
Eq. (\ref{Omega}).  Therefore, when $D\leq D_{br}$ or $f\geq f_{br}$
the shear from induced quadrupole moment is strong enough to break the
crust and beyond this point a solid crust can no longer exist.  This
frequency is below the observable region from $100$ to $1000$ Hz for
current and proposed gravitational wave detectors such as LIGO
\citep{LIGO}.  Consequently, during the last stages of inspiral that
are observed in gravitational waves, effects stemming from the solid
crust are probably irrelevant and our calculations assuming a liquid
phase should be valid.

\begin{figure}[tb]
	\includegraphics[width=8.8cm, angle=90]{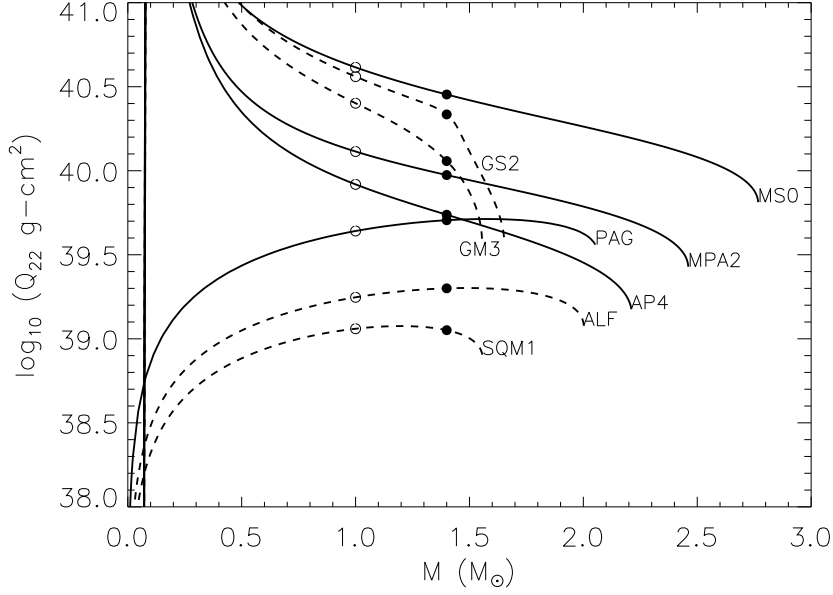}
	\caption{The maximum quadrupole moment $Q_{22,max}$ of a solid
          crust as a function of mass for normal and strange matter
          stars. Filled (open)
          circles indicate configurations with $M=1.4$ M$_\odot$ (1.0
          M$_\odot$).}
	\label{Q22eosNS}
\end{figure}

Using the expressions provided by \citet{Owen2005}, which are supported by
our results, we can approximate
the maximum quadrupole moment for a solid crust through
\begin{equation}
Q_{22,max}=\frac{\sigma_{max}}{0.01}\,\rm{g\,cm}^2\left\{
\begin{array}{ll}
2.4\times\, 10^{38}\left(\frac{R}{10\rm{\,km}}\right)^{6.26} \left(\frac{1.4 \, M_\odot}{M}\right)^{1.2} & \text{neutron stars,}\\
3.5\times\, 10^{39}\left(\frac{R}{8\rm{\,km}}\right)^{6} \left(\frac{1.4 \, M_\odot}{M}\right) & \text{hybrid and meson-condensate stars,}\\
2.8\times\, 10^{41}\frac{\mu}{4 \, 10^{32}\rm{\,erg/cm}^3}\left(\frac{R}{10\rm{\,km}}\right)^{6} \left(\frac{1.4 \, M_\odot}{M}\right) & \text{solid strange stars,}
\end{array}\right.,
\end{equation}
where $\sigma_{max}=0.1$ is the breaking strain of the crust and
$\mu \approx 4 \times 10^{32}\rm{\,erg/cm}^3$ is a typical shear
modulus of a strange quark matter crust (\citet{Horowitz2009a}), which is
a thousand times the typical value in the crust of a normal neutron
star.  The results are shown in Fig. \ref{Q22eosNS}. For stars with
masses heavier than $1 M_\odot$ the maximum quadrupole moments are
within an order of magnitude of the typical value of $10^{40}
{\rm{\,g\,cm^2}}$.

\begin{figure}[tb]
	\includegraphics[width=8.8cm,angle=90]{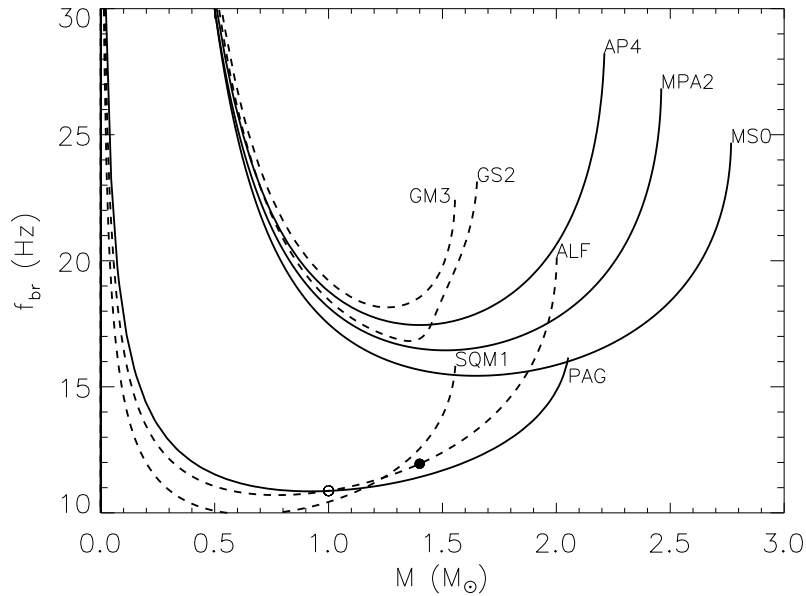}
	\caption{The frequency of gravitational waves from an
          inspiraling binary when tidal forces are expected to break
          the crust for normal and strange quark matter stars. Filled (open)
          circles indicate configurations with $M=1.4$ M$_\odot$ (1.0
          M$_\odot$).}
	\label{fbreosNS}
\end{figure}

Fig. \ref{fbreosNS} shows results for the breaking frequency $f_{br}$
calculated utilizing Eq. (\ref{gwf}) with the appropriate values for
$Q_{22,max}$ from Fig. \ref{Q22eosNS}.  The breaking frequency for
both kinds of stars heavier than a few tenths of a solar mass is well
below the LIGO lower boundary of $100$ Hz~\citep{LIGO}.
Therefore, the crust may
be assumed to be melted during the time it is observed, and the
approximation of treating the entire star as a liquid is justified. 

\section{Summary and Conclusions}

The quadrupole polarizabilities of normal neutron stars and self-bound
quark matter stars have been calculated for a wide class of proposed
equations of state of dense matter for both normal and strange quark
matter stars. The quadrupole polarizabilities $\lambda=2 R^5 k_2/(3 \,
G)$ are characterized by the dimensionless Love number $k_2$ and both
are sensitive to the equation of state, in particular to the
compactness parameter $M/R$ and the overall compressibility of the
equation of state. For normal neutron stars, $k_2$ and $\lambda$
exhibit pronounced maxima for configurations with masses close to a
solar mass for most equations of state. The maximum value of $k_2$ is
not very sensitive to the EOS, lying in the range 0.1--0.14.  In each
case, maximum mass configurations have significantly lower values of
$k_2$ and $\lambda$ than their solar mass counterparts.

Love numbers for self-bound strange quark matter stars with or without
crusts are qualitatively different than those of normal neutron
stars. The maxima in the value of $k_2$ for strange quark matter stars
without crusts occurs for masses less than 0.1 M$_\odot$, and maximum
values of order 0.8 are achieved.  As in the normal matter case, the
maxima in quadrupole polarizabilities occurs for configurations near 1
M$_\odot$.  In contrast, the magnitudes of quadrupole polarizabilities
of strange quark matter stars are usually much less than those of
normal stars, owing to the larger radii of the latter.  

Our investigations also point the need to examine the core-crust
interface region of both normal and self-bound quark matter stars more
closely.  The important issue that bears close scrutiny is the precise
nature (first or second order) of possible phase transitions.  In the
case that strong discontinuities exist near the core-crust interface
of strange quark matter stars, dimensionless Love numbers are
suppressed for low mass stars relative to the cases for which there is
no crust.  However, for stars of order 1 M$_\odot$ or larger, the
presence or absence of a crust has
little influence on Love numbers.

The strength of the tidal signatures from coalescing compact binaries
is proportional to $\lambda$, and is therefore quite sensitive to the
radii of the stars.  For stellar configuratons with radii of order 11
km or less, the tidal response might be too small to observe, implying
that a positive detection might be sufficient to rule out the presence
of a self-bound star, such as a strange quark matter star, in the
observed system.
\acknowledgements 
The authors thank Ben Owen for alerting them to Love's importance.
Thanks are also due to Dany Page, Andrew Steiner and Mark Alford for
providing equations of state based on their work.  SP and MP
acknowledge research support from the U.S. DOE grant
DE-FG02-93ER-40756.  JML acknowledges research support from the
U. S. DOE grant DE-AC02-87ER40317 and from a Glidden Visiting
Professorship Award at Ohio University.
\appendix
\section{Parameters for Analytic Solutions of Einstein's Equations}

We use the notation $\beta=GM/Rc^2, ~\alpha=4\pi\rho_c R^2$ and $x=(r/R)^2$.

\subsection{Uniform Density ($\rho=\rho_c$)}

\begin{eqnarray}
&&\alpha=3\beta,  \qquad e^{-\lambda}=1-2\beta x,
\nonumber\\
&&{p\over\rho_c}={\sqrt{1-2\beta}-\sqrt{1-2\beta x}
\over\sqrt{1-2\beta x}-3\sqrt{1-2\beta}},\qquad
c_s^2=\infty\,,\qquad {\rho_s\over\rho_c}=1\,.
\label{eq:inc}
\end{eqnarray}
\subsection{Tolman VII ($\rho=\rho_c[1-x]$)~\citep{To1939}}

\begin{eqnarray}
&&\alpha={15\over2}\beta,\qquad e^{-\lambda}=1-\beta x(5-3x)\nonumber\\
&&{p\over\rho_c}={2\over15}\sqrt{3\over\beta
    e^\lambda}\tan\phi-{1\over3} + {x\over5},\nonumber\\
&&\phi={w_1-w\over2}+\phi_1, \qquad\phi_1=\tan^{-1}
\sqrt{\beta\over3(1-2\beta)},\nonumber\\
&&w=\ln\left[x-{5\over6}+\sqrt{e^{-\lambda}\over3\beta}\right],\qquad
w_1=\ln\left[{1\over6}+\sqrt{1-2\beta\over3\beta}\right],\nonumber\\
&&c_s^2={\tan\phi\over5}\left[\tan\phi+\sqrt{\beta\over3e^\lambda}
(5-6x)\right]\,.
\label{eq:tol7}
\end{eqnarray}

\subsection{Buchdahl's Solution ($\rho=12\sqrt{p_*p}-5p$)~\citep{Bu1959,Bu1966}} 

\begin{eqnarray}
&&\alpha=\pi^2\beta
(1-\beta)^2{1-5\beta/2\over1-2\beta},\qquad 
z={1-\beta\over1-\beta+u}\pi\sqrt{x}\,,\nonumber\\
&&u=\beta{\sin z\over z},\qquad
e^\lambda={(1-2\beta)(1-\beta+u)\over(1-\beta-u)
\left(1-\beta+\beta\cos z\right)^2},\qquad
c_s^2={u\over1-\beta-4u}\,,\nonumber\\
&&{\rho\over\rho_c}={(1-2\beta)(2-2\beta-3u)\over(2-5\beta)(1-\beta+u)^2}{u\over\beta},\qquad
{p\over\rho_c}={\beta(1-2\beta)\over(1-\beta+u)^2(2-5\beta)}\left({u\over\beta}\right)^2\,.
\end{eqnarray}

\subsection{Generalized Tolman IV (N=1)~\citep{Na1950,Na1951,La2007}}

\begin{eqnarray}
&&\alpha={3\beta\over2}{2-3\beta\over1-3\beta},\qquad 
e^\lambda={1-3\beta+2\beta x\over(1-3\beta+\beta x)(1-\beta x)},\nonumber\\
&&{\rho\over\rho_c}={1-3\beta\over2-3\beta}{(2-3\beta)
(1-3\beta)+\beta(3-7\beta)x+2\beta^2x^2\over(1-3\beta+2\beta x)^2}\,,\quad
{\rho_s\over\rho_c}={(1-2\beta)(1-3\beta)\over(1-3\beta/2)(1-\beta)}
\,,\nonumber\\
&&{p\over\rho_c}={1-3\beta\over2-3\beta}\beta{1-x\over1-3\beta+2\beta x},
\qquad c_s^2={1-3\beta+2\beta x\over5-15\beta+2\beta x}\,.
\end{eqnarray}
\subsection{Generalized Tolman IV (N=2)~\citep{Na1950,Na1951,La2007}}

\begin{eqnarray}
&&\alpha=3\beta\left({2-2\beta\over2-5\beta}\right)^{2/3},\qquad
e^{-\lambda}=1-2\left({2-2\beta\over2-5\beta+3\beta x}\right)^{2/3}
\beta x\,,\nonumber\\
&&{\rho\over\rho_c}=\left(1+{5\beta x\over3(2-5\beta)}\right)
\left(1+{3\beta x\over2-5\beta}\right)^{-5/3}\,,\quad
{\rho_s\over\rho_c}={(1-5\beta/3)(1-5\beta/2)^{2/3}\over(1-\beta)^{5/3}}
\nonumber\\
&&{p\over\rho_c}=\left({2-5\beta\over2-2\beta}\right)^{2/3}
{1\over3(2-5\beta+\beta x)}\left[2-\left({2-2\beta\over2-5\beta+3\beta x}
\right)^{2/3}(2-5\beta+5\beta x)\right],\nonumber\\
&&c_s^2={2-5\beta+3\beta x\over5(2-5\beta+\beta x)^3}
\left[{(2-5\beta+3\beta x)^{5/3}\over(2-2\beta)^{2/3}}+
(2-5\beta)^2-5\beta^2x^2\right]\,.
\end{eqnarray}

\bibliographystyle{apsrev}
\bibliography{refs}

\end{document}